\documentclass[prl,twocolumn,aps,showpacs]{revtex4}

\usepackage{latexsym}
\usepackage{graphicx}
\usepackage[english]{babel}

\begin{document}

\author{A. Gomez-Marin$^{1}$, J. Garcia-Ojalvo$^{2}$ and J. M. Sancho$^{1}$}
\title {Self-sustained spatiotemporal oscillations induced by
membrane-bulk coupling}
\affiliation{$^{1}$Facultat de Fisica, Universitat de Barcelona, Diagonal 647, 08028 Barcelona, Spain. \\
$^{2}$Departament de Fisica i Enginyeria Nuclear, Universitat Politecnica de Catalunya, Colom 11, E-08222 Terrassa, Spain.} 

\date{\today}

\begin{abstract}
We propose a novel mechanism leading to spatiotemporal oscillations in
extended systems that does not rely on local bulk instabilities. Instead,
oscillations arise from the interaction of two subsystems of different spatial
dimensionality. Specifically, we show that coupling a passive
diffusive bulk of dimension $d$ with an excitable membrane of
dimension $d-1$ produces a
self-sustained oscillatory behavior.
An analytical explanation of the phenomenon is provided for $d=1$.
Moreover, in-phase and anti-phase synchronization of oscillations are found numerically in one and two dimensions. This novel dynamic instability could
be used by biological systems such as cells, where the
dynamics on the cellular membrane is necessarily different from that of
the cytoplasmic bulk.
\end{abstract}


\maketitle

\emph{Introduction.} In spatially extended systems,
spatiotemporal oscillations usually arise from
short-wavelength, finite-frequency instabilities that affect the local dynamics of the system's bulk. Within that scenario, boundaries are reduced to passive
elements that play somewhat secondary roles, such as wavelength
discretization and wavevector selection \cite{cross93}.
There are many situations in nature, however, where boundaries
have {\em active} dynamics. Fronts are known to be initiated, for instance,
at the interface between different catalytic components in
microcomposite surfaces \cite{imbihl}. Similarly, chaotic dynamics has been
shown to arise in a catalytic surface coupled to a (passive) gas phase
\cite{pt}. Other examples of active surfaces include Langmuir monolayers
\cite{langmuir} and membranes with active proteins such as
proton pumps \cite{prost}. These systems delimit regions of
higher dimensionality, which usually have different dynamics from that
of the active boundary.

Few studies have addressed the interplay between different dynamics of
a bulk and its boundary. In \cite{levine} it has shown that an active membrane can give
rise to stationary cytoplasmic patterns.
Special attention has been paid to the issue
of pole-to-pole protein oscillations underlying symmetrical
cell division in bacteria \cite{meinhardt, howard05}.
Most models of this phenomenon assume that the main
source of the oscillations are biochemical reactions occurring
at the cell membrane, with the cytoplasm being mainly passive,
hosting at most phosphorylation reactions
\cite{howard01,kruse02,hwang,kulkarni,drew}. It is thus of our interest to determine whether nontrivial dynamics can arise in the limiting case of an active boundary delimiting a purely passive bulk. 

Motivated by
this system, which is known to include activator and inhibitor proteins
\cite{meinhardt}, and taking into account the fact that activator-inhibitor
dynamics sustains excitability \cite{mikhailov}, we will consider in what
follows an excitable membrane, limiting an otherwise purely passive
bulk \cite{pando}.
Indeed, we will show that this simplified scenario is able to sustain {\it dynamic} spatiotemporal oscillations in a wide
parameter range, even though neither the bulk nor the boundary are
oscillatory. 
Moreover, we will provide an analytical explanation for this effect in
the case of a one-dimensional bulk and point-like boundaries. This
analysis allows us to predict the extent of the oscillatory region in terms of the relevant parameters: the system length and the  coupling strength between the bulk
and the boundaries.

\emph{The model.} We consider a spatially-extended passive system
affected by simple diffusion and linear degradation, bounded by
an active membrane with activator-inhibitor dynamics.
The equations that mathematically describe the bulk are
\begin{eqnarray}
\label{bulkU}
\partial_{t}U=D_{U}\nabla^{2}U -\sigma_U U,
\\
\label{bulkV}
\partial_{t}V=D_{V}\nabla^{2}V-\sigma_V V ,
\end{eqnarray}
where $U$ and $V$ are the concentrations of activator and inhibitor,
respectively, the corresponding diffusion coefficients are $D_U$ and
$D_V$, and both species are assumed to decay at rates $\sigma_U$
and $\sigma_V$  \cite{comment}. The results shown below do not change in the presence of a constitutive expression of the bulk species, represented by the addition of constant terms in the r.h.s. of equations~(\ref{bulkU})-(\ref{bulkV}).

The dynamics at the system's boundary is given by
\begin{eqnarray}
\label{membu}
\dot{u}=f(u,v) + k_{u} (\hat{n} \cdot  \vec{\nabla}U) ,
\\
\label{membv}
\dot{v}=\epsilon g(u,v) + k_{v} (\hat{n} \cdot \vec{\nabla} V) .
\end{eqnarray}
The reaction terms $f(u,v)$ and $g(u,v)$ are chosen to account for a
local activator-inhibitor dynamics, so that the $u$-nulcline [$f(u,v)=0$]
has a cubic shape in the $(u,v)$ space, while the $v$-nulcline
[$g(u,v)=0$] is monotonically increasing, as in typical Fitzhugh-Nagumo models (see Fig.~\ref{nul} below). 
In particular, we have considered the following expression in dimensionless form
\begin{eqnarray}
\label{fuv}
&&f(u,v)=u-q(u-2)^{3}+4-v,
\\
\label{guv}
&&g(u,v)=u z-v,
\end{eqnarray}
where $z$ and $q$ control the shape of the nulclines. The parameter
$\epsilon$  that accompanies $g(u,v)$ determines a different time scale
for both species, so that $u$ is much faster than $v$ if $\epsilon\ll1$.
We consider in what follows $z=3.5$ and $q=5$, which renders the
membrane excitable. Under this condition, the membrane (when isolated
from the bulk) is at rest, and only when a small perturbation is applied to
it, an excursion corresponding to an activator pulse is produced.
The second term in the right-hand-side of  Eqs.~(\ref{membu}) and
(\ref{membv}) accounts for the exchange of species between the
membrane and the bulk.
The constants $k_u$ and $k_v$ determine the coupling strength,
while $\hat{n} \cdot \vec{\nabla} C(\vec{r})$ is the normal derivative of the concentration field $C(\vec{r})$, with $\hat{n}$ being a unit vector
normal to the boundary and pointing towards the bulk.

\emph{Phenomenology.} In principle, one would expect the system
described above to be quiescent unless a perturbation is applied
to the membrane. Such a perturbation would excite a concentration
pulse at the membrane, which would then propagate into the
diffusive bulk and progressively decay. However, coupling between
the membrane and the bulk can give rise to interesting new phenomenology.
Due to the coupling, degradation of the inhibitor in the bulk leads to
a decrease in the concentration of the inhibitor also in the membrane,
which subsequently allows, via the activator-inhibitor dynamics, a pulse
in the activator concentration. After some time, the excitability
mechanism reduces the activator level spontaneously back to the resting
state, and the process can restart again and repeat endlessly, leading to
self-sustained oscillations even when neither the membrane nor (evidently)
the bulk are intrinsically oscillatory. This is indeed observed in our
model, as we show below.

\begin{figure}[htb]
\begin{center}
  \includegraphics[width=0.4\textwidth]{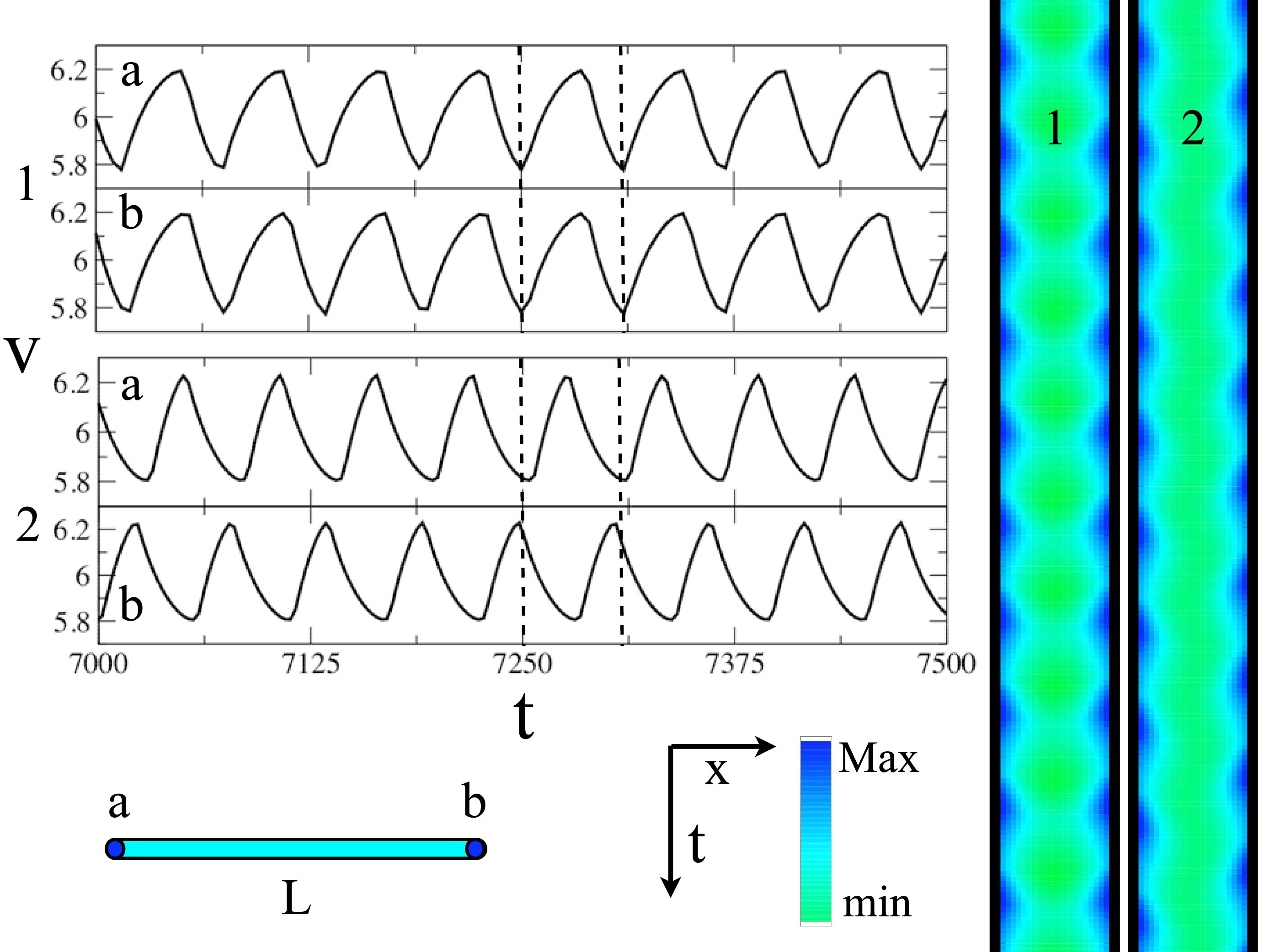}
  \caption{Oscillations of the membrane inhibitor $v$ in a one-dimensional
bulk, bounded by two 0-dimensional ``membranes''. The left panel
shows time series in each of the two boundaries (denoted respectively
by a and b). Two regimes are shown: in-phase and anti-phase oscillations,
labeled ``1'' and ``2'', and  are obtained by setting 
$D_V=0.7$ and $D_V=0.3$, respectively. 
The two right panels show
spatiotemporal representations of the inhibitor concentration in color code, with time running
from top to bottom and space represented horizontally.
The parameters used in these simulations are $L=10$,
$\epsilon=0.015$, $k_v=5\epsilon$, $\sigma_V=D_V/100$ and $\sigma_U=D_U=k_u=0$. As initial condition, the system
is set at the rest state with additional (small) spatial fluctuations, 
so that there is a slight heterogeneity in the initial concentration field.  
}
  \label{fig1}
\end{center}
\end{figure}

Once oscillations are autonomously occurring, we can analyze what
happens when the bulk is limited by two opposing boundaries, similarly to an ellipsoidal bacterial cell.
If the distance between the poles is small
enough the oscillations interfere, and eventually become
synchronized. This is shown in the one-dimensional numerical simulations
presented in Fig.~\ref{fig1}. Two regimes corresponding to in-phase
and anti-phase oscillations of the two poles are observed.
For each regime, a plot of the time evolution of $v$ at the two
boundaries is displayed. Additionally, spatiotemporal plots are shown in
the right panels. The
results clearly show
not only that self-sustained oscillations appear even when the system is not intrinsically oscillatory, but that the oscillations can synchronize either
in phase or in anti-phase. The behavior persists even when the system
is perturbed by noise (data not shown),  evidencing the robustness of the phenomenon.
In Fig.~\ref{fig1}, the inhibitor diffusion was varied in order to change the
type of phase locking. As we show below, other parameters can be similarly tuned to control the system's behavior.

\emph{Theoretical analysis in $d=1$.}  The effect of coupling on the behavior of
the excitable membrane can be determined by using the solution and
boundary conditions 
of the bulk equations (\ref{bulkU})-(\ref{bulkV}) in the membrane
equations (\ref{membu})-(\ref{membv}). To that end, we determine the
stationary solutions of the diffusion
equations (\ref{bulkU})-(\ref{bulkV}) in a one-dimensional region of length
$L$. In the case of the inhibitor, the resulting density profile is
$V(x)=V(0) \cosh \left[\sqrt{\frac{\sigma_V}{D_V}}  (\frac{L}{2}-x) \right]/ 
 \cosh \left[\sqrt{\frac{\sigma_V}{D_V}} \frac{L}{2} \right] $, where $V(0)=v$. 
Its derivative at the boundary is then
\begin{equation}
V'(x=0)=-v  \sqrt{\frac{\sigma_V}{D_V}} \tanh \left[\sqrt{\frac{\sigma_V}{D_V}} 
  \frac{L}{2} \right].
\end{equation}
This expression, when
inserted into the steady-state membrane equation  
\begin{equation} \label{vsst}
0=\epsilon  g(u,v) + k_v V'(x=0),
\end{equation}
leads to a new effective inhibitor nulcline  
\begin{equation} \label{1dnulc}
v=\frac{z \; u}{1+ \frac{l_{1}}{l_{0}}  \tanh
\left[ \frac{L/2}{l_{0}} \right]}. 
\end{equation}
Here  $ l_{1} \equiv \frac{k_v}{\epsilon}$  and $l_{0} \equiv \sqrt{\frac{D_V}{\sigma_V}}$.
Note that the initial inhibitor nulcline $v=zu$ is modified due to the coupling with the bulk, effectively decreasing its slope $z$ as the coupling increases.
The same analysis can be done for the
activator, although in that case, the contribution from the coupling with the bulk barely modifies the
shape of the membrane $u$-nulcline [$f(u,v)=0)$], and therefore its
effects will be ignored in what
follows. 

\begin{figure} [t]
\begin{center}
\includegraphics[width=0.4\textwidth]{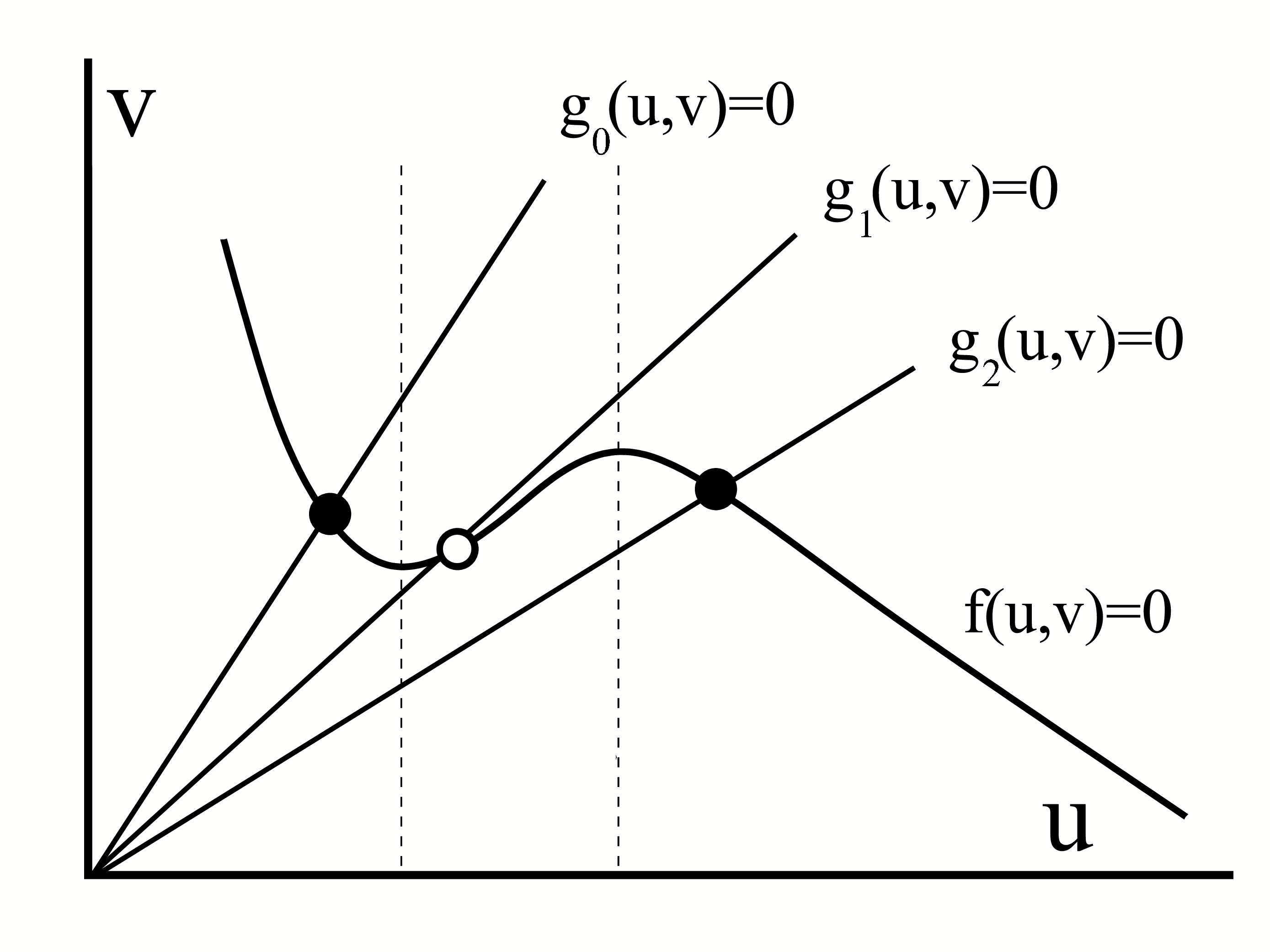}
\caption{Qualitative scheme of the effect of the bulk on the nulclines of the excitable membrane.
A full circle denotes a stable fixed point and an empty circle an
unstable one.
The $g$ nulcline labeled $g_0$ corresponds to the absence of
coupling with the bulk. $g_1$ and $g_2$  define modified $v$-nulclines for increasing $k_v \neq 0$. The case of $g_1$ corresponds to a self-sustained oscillatory regime, while for larger coupling ($g_2$) the system is
back at a steady state again. The $f$ nulcline is practically unmodified by the coupling, and is left unchanged in this schematic plot. 
}
\label{nul}
\end{center}
\end{figure}

With the above considerations, we can now understand, both
qualitatively and quantitatively, the effect of membrane-bulk coupling.
As shown in Fig.~\ref{nul}, the effect of the diffusive and degrading
bulk can be mapped to an effective variation of the slope of the $g(u,v)$ nulcline, yielding a change from an excitable situation where only
one stable fixed point exists, to an oscillatory regime where the fixed
point is unstable and a limit cycle (not shown) develops. Finally, large
enough coupling even leads the system back again to an excitable regime.
The result obtained in Eq.~(\ref{1dnulc}) shows that three length scales
control the system's behavior in the steady state: $L$, $l_0$ and $l_1$.
The first one, $L$, is the natural length of the system. 
The second one, $l_0$, is a characteristic length determined by the 
ratio of the diffusion $D_V$ and the degradation $\sigma_V$, and corresponds
to an action length of the bulk. Finally, the effect of the
coupling enters directly through the
combination of the coupling strength $k_v$ and the time scale ratio
$\epsilon$, giving rise to a characteristic length scale $l_{1}$. 

\begin{figure}[t] 
\begin{center}
\includegraphics[width=0.4\textwidth]{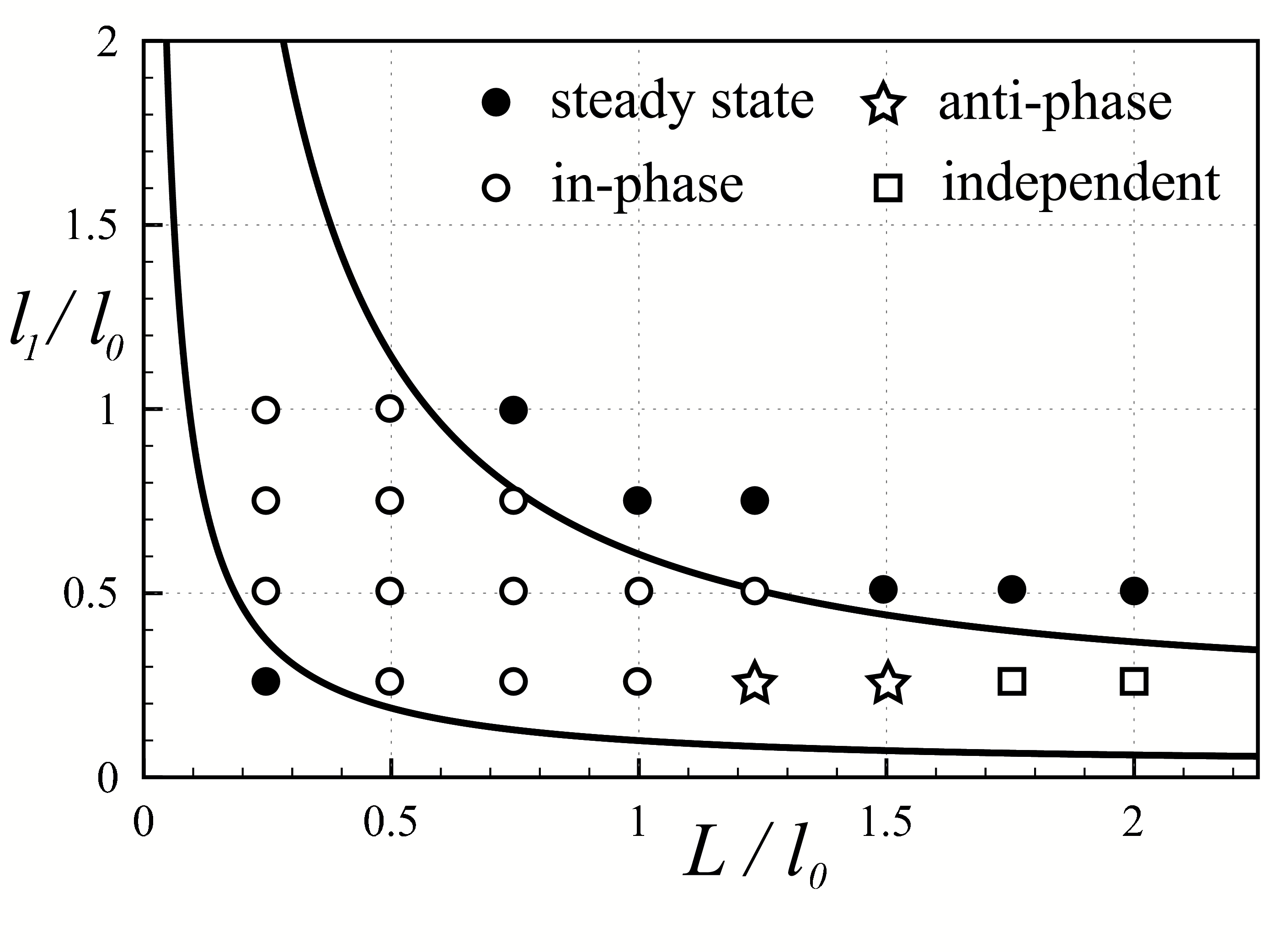}
\caption{Phase diagram of a one-dimensional diffusive and degrading
bulk delimited by
two point excitable boundaries. Solid lines have been determined
analytically. The oscillatory region lays within these lines. 
Symbols correspond to numerical simulations: full circles, stationary state; empty circles, in-phase oscillations; empty stars, anti-phase oscillations; empty squares, independent oscillations.
Parameters are those of Fig. \ref{fig1} except for $D_V = 0.5$, and $k_v$ and $L$ are varied.} 
\label{phased}
\end{center}
\end{figure}

Making use of Eq.~(\ref{1dnulc}) and considering the location and 
the stability of the fixed point (determined by the crossing of the nulclines), we can analytically predict the
region in parameter space where the system will behave
as an oscillator. The corresponding phase diagram is shown in
Fig.~\ref{phased}.
The theoretically predicted boundaries of the oscillatory region
(solid lines in Fig.~\ref{phased}) agree
very well with simulations. 
Therefore, the steady state analysis yields
a clear understanding of why the excitable membrane becomes unstable and starts to oscillate. 
The second set of important properties of the system, i.e. synchronization
and phase locking between the oscillating poles, have to be determined
numerically. 
Such an investigation leads to different types of oscillatory
regimes, as shown in Fig.~\ref{phased}, in which anti-phase oscillations (stars)
separate the region of in-phase oscillations (empty circles) from a domain where the two poles of the system oscillate independently (squares), namely, having the same frequency but an undefined phase relation.
Note that for large $L$ the pole oscillations
can not interact, since the opposing membranes are too far away.
On the other hand, for large enough $k_v$ (i.e. $l_1$) oscillations (when they occur) can only be in phase. Finally, we note that increasing the inhibitor diffusion coefficient $D_V$ makes $l_0$ shorter, which changes the overall scale of the phase diagram.

\emph{Extension to $d=2$: numerical simulations.} Once the one-dimensional
case is understood and fully characterized, we now consider a
two-dimensional system with a rectangular geometry.
The main difference is that now the membrane is also distributed
in space (but without internal diffusion \cite{howard01}), while in the previous case
it was zero-dimensional.
However, by analogy to the one-dimensional case the relevance of the
different parameters can be understood, and again the same phenomena are encountered. Every parameter has an important physical meaning.  $D_{V}$ is responsible for the indirect coupling between membrane elements and
is thus a necessary ingredient for the synchronization.
$D_{U}$ is  crucial to avoid several pathologies that occur in $d>1$, such as accumulation of inhibitor along the membrane, which leads to
local blocking of the oscillations. $\sigma_V$ is
key to the spontaneous onset of local oscillations at the membrane, as we
have already seen in the one-dimensional case. 
$k_{v}$ determines the new effective nulcline due to the influence
of the bulk, and controls the synchronization phases.

Furthermore, there are now two length scales in the problem associated with the system size.
$L_1$ is the length of the rectangular system and its width is denoted by $L_2$. Assuming
$L_1>L_2$, we are interested in observing oscillations along 
$L_1$, mimicking the pole-to-pole oscillations in elongated bacteria.
In this case, the lateral walls can dominate the main activity of the system, 
preventing anti-phase
oscillations for short lengths. For very long lengths the communication is
fragile.  For intermediate lengths $L_1$, given that $L_2$ is smaller
than $L_1$, we expect the
oscillations along the minor axis to be
more or less synchronized in phase, and not to
disturb excessively the pole-to-pole oscillations.

\begin{figure}[t]
\begin{center}
\includegraphics[width=0.4\textwidth]{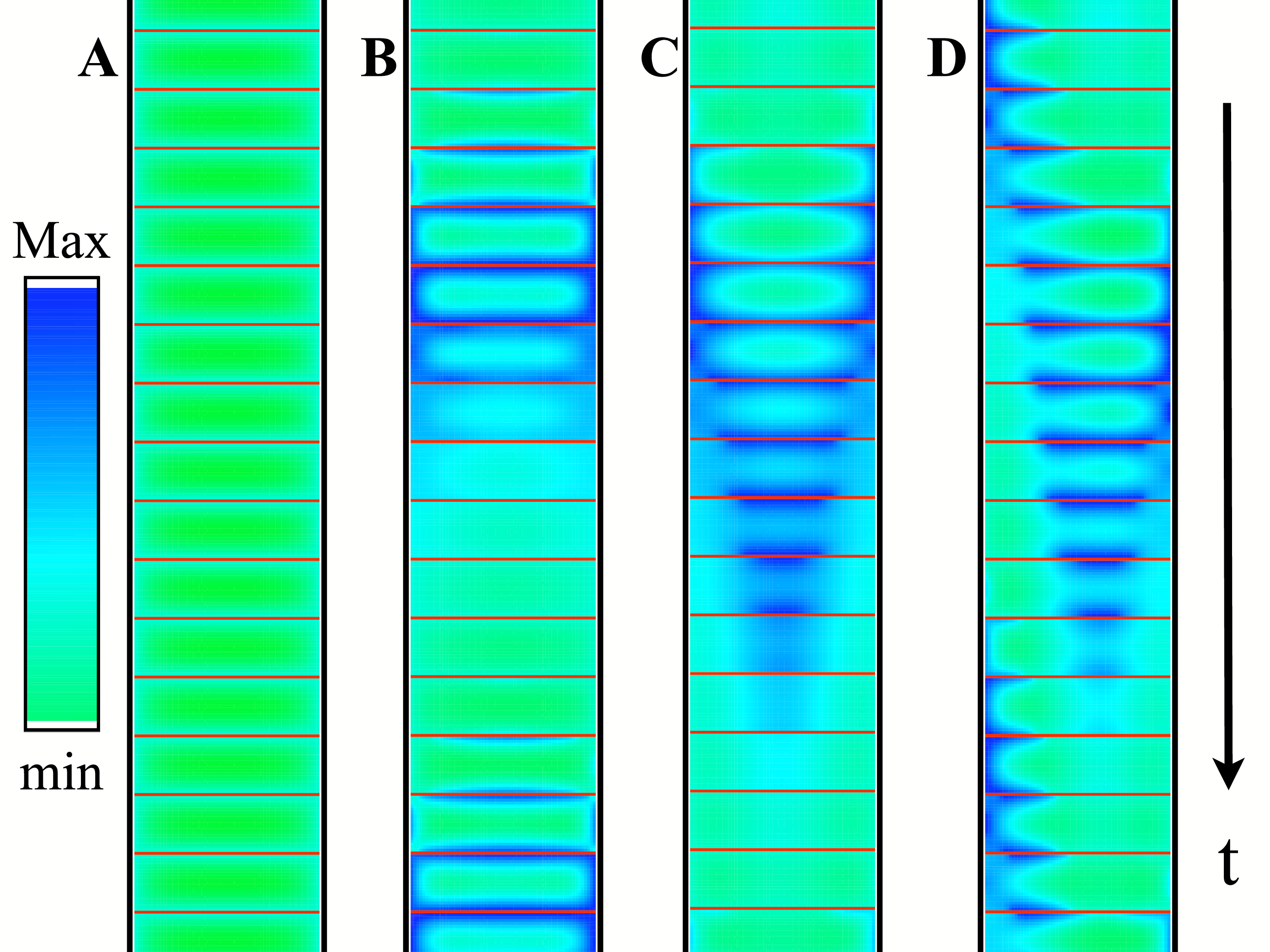}
\caption{Temporal series of snapshots of $V$ in 2-dimensions. {\bf A}. Steady state solution. {\bf B} and {\bf C}. In-phase periodic oscillations. {\bf D}. Anti-phase periodic oscillations and traveling wave. 
The values of the parameters are $L_1=15.5$, $L_1/L_2=3.4$, $D_U=D_V=0.1$, $\sigma_V=0.00124$, $\sigma_U=10^{-4}\sigma_V $,  $k_u=k_v \equiv k$ and $\epsilon=0.015$.
The different regimes are obtained varying $k$: ${\bf A}$, $k=0.03$; ${\bf B}$, $k=0.08$; ${\bf C}$, $k=0.095$; and
${\bf D}$, $k=0.11$. Each snapshot is taken every $\Delta t=10$.}
\label{fig4}
\end{center}
\end{figure}
We now present the phenomenology that can be observed numerically
in two dimensions. 
In Fig.~\ref{fig4}, temporal sequences of two-dimensional snapshots of the density map of $V$ are displayed for different kinds of behavior.
The system is initiated again from the rest state, superimposed with 
small random heterogeneous perturbations.
When dynamical noise is added to the simulations, not
only is synchronization maintained,
but also the different regimes develop easier and faster.
Figure~\ref{fig4}A corresponds to small couplings, for which
the system is unable to oscillate and remains in a stable fixed
point, corresponding to the quiescent state of the excitable membranes.
For larger coupling (panel B), in-phase oscillations emerge from the
center of the system. Panel C shows  
a different kind of in-phase oscillation, which is governed by the poles
of the system. Finally, panel D shows a traveling
wave that leads to anti-phase oscillations of the poles, similarly to
what was found in the one-dimensional case.
The traveling wave alternates periodically and slowly its direction of
motion, from
left to right and vice versa (with a period larger than the time span
shown in Fig.~\ref{fig4}). In spite of these alternations, the poles 
oscillate periodically and in perfect anti-phase.  The
dimensionless parameters leading to this regime, given in
the caption of Fig.~\ref{fig4}, correspond to reasonable biological values when turned into dimensional units. For instance, just by assuming a bacterium of length $L\simeq 4 \mu m$, it is found that $l_0\simeq 2.28 \mu m$ and $l_1=1.88\mu m$.

\emph{Conclusions.}
We have reported a minimal mechanism that generates pole-to-pole
oscillations in non-active elongated media. The mechanism relies on
the interaction of the system's bulk with an excitable (non-oscillatory)
membrane, and can be understood analytically for one-dimensional
bulks making use of a phase-plane
picture of the membrane's excitability. Both in-phase and anti-phase
oscillations
can be observed. In the case of a two-dimensional bulk, anti-phase
dynamics is associated with a traveling wave that periodically reverses
its propagation direction. This could be a generic mechanism leading
to spatiotemporal oscillations in systems limited by active boundaries,
such as cells.

This research was supported by Ministerio de Educacion y Ciencia (Spain)
under project FIS2006-11452 and grant  FPU-AP-2004-0770 (A.G-M), and
by the Generalitat de Catalunya.

\end{document}